\documentclass[prb,aps,preprint]{revtex4}
\newcommand{\beq}{\begin{equation}}
\newcommand{\eeq}{\end{equation}}
\usepackage{graphics}
\usepackage{epsf}
\usepackage{epsfig}
\usepackage{amsmath}
\usepackage{amsfonts}
\usepackage{amssymb}
\usepackage{graphicx}
\usepackage{epsfig}

\begin{document}
\title{Glass-like ordering and spatial inhomogeneity of magnetic structure in Ba$_3$FeRu$_2$O$_9$ : The role of Fe/Ru-site disorder}
\author {Srimanta Middey$^1$ and Sugata Ray$^{1,2,\star}$}
\affiliation {$^1$Centre for Advanced Materials,~Indian Association for the Cultivation of Science, Jadavpur, Kolkata 700032, India}
\affiliation {$^2$Department of Materials Science, Indian Association for the Cultivation of Science, Jadavpur, Kolkata 700032, India}
\author{K. Mukherjee, P. L. Paulose, and E. V. Sampathkumaran}
\affiliation{Tata Institute of Fundamental Research, Homi Bhabha Road, Colaba, Mumbai 400005, India}
\author{C. Meneghini}
\affiliation{Dipartimento di Fisica, Universita´ di ''Roma Tre'', Via della vasca navale, 84 I-00146 Roma, Italy}
\author{S. D. Kaushik and V. Siruguri}
\affiliation{UGC-DAE-Consortium for Scientific Research Mumbai Centre, R5 Shed, Bhabha Atomic Research Centre, Mumbai 400085, India}
\author{Kirill Kovnir}
\affiliation{Department of Chemistry and Biochemistry, Florida State
University, Tallahassee FL, USA}
\author{D. D. Sarma}
\affiliation{Solid State and Structural Chemistry Unit, Indian Institute of Science, Bangalore-560012, India}

\date{\today}

\begin{abstract}
Several doped 6$H$ hexagonal ruthenates, having the general formula Ba$_{3}M$Ru$_{2}$O$_{9}$, have been studied over a significant period of time in order to understand the unusual magnetism of ruthenium metal. However, among
them, the $M$=Fe compound appears different since it is observed that unlike others, the 3$d$ Fe ions and 4$d$ Ru
ions can easily exchange their crystallographic positions and as a
result many possible magnetic interactions become realizable. The
present study involving several experimental methods on this
compound establish that the magnetic structure of
Ba$_{3}$FeRu$_{2}$O$_{9}$ is indeed very different from all other
6$H$ ruthenates. Local structural study reveals that the possible
Fe/Ru-site disorder further extends to create local chemical inhomogeneity,
affecting the high temperature magnetism of this
material. There is a gradual decrease of $^{57}$Fe
M\"{o}ssbauer spectral intensity with decreasing temperature (below
100 K), which reveals that there is a large spread in the
magnetic ordering temperatures, corresponding to many spatially inhomogeneous regions. However, finally at about 25 K, the whole compound is found to take up a global glass-like magnetic ordering.
\end{abstract}

\maketitle

PACS number(s): 75.50.Lk, 74.62.En, 78.30.Ly


\section{Introduction}

Ruthenium based oxides have been drawing attention of the community
for quite some time because they are the most salient examples of
unusual 4$d$ magnetism. One of the most well discussed compounds
among the ruthenates in this regard is SrRuO$_3$,~\cite{srruo3_prb}
which has an orthorhombic structure (Space group: $Pnma$) and
exhibits ferromagnetism below a Curie temperature ($T_c$) of 160~K.
Interestingly, the analogous CaRuO$_3$ does not order magnetically
down to the lowest temperature~\cite{caruo3_para_PRB} even though
the crystal structure is similar to SrRuO$_3$. However, structurally
 BaRuO$_3$ is an exotic system, which is known
to adopt at least four different crystal structures depending on the
synthesis conditions.~\cite{baruo3_PNAS} Among these four
structures, the most recently reported cubic polymorph stabilizes
only at a very high pressure and shows ferromagnetic ordering below
60~K.~\cite{baruo3_PNAS} The other three polymorphs are all
hexagonal, having differences in octahedral connectivities along the
$c$-axis, and all of them exhibit paramagnetic
behavior.~\cite{Cava_PRB,zhao_JSSC} The common factor between these
three hexagonal polymorphs is the existance of two different
octahedral sites for Ru ions in the crystal structure, giving rise
to simultaneous presence of Ru-Ru and Ru-O-Ru connectivities. As a
result, this particular structural geometry offers a unique
opportunity to introduce different metal ions in one of these two
octahedral sites and to probe the onset of various possible magnetic
interactions involving ruthenium. Based on this idea, the structure and magnetic properties of different metal ruthenates
Ba$_3$$M$Ru$_2$O$_9$ ($\frac{1}{3}$rd Ru replaced by $M$ in BaRuO$_3$) were scrutinized, where $M$ is varied from transition metals
{\it e.g.} Ti, Mn, Fe, Co, Ni, Cu, Zn, and
Cd,~\cite{verdoes_AC,Schaller_MRB,zhao_JSSC1,donohue_IC,Battle_JSSC,Cava_JSSC,Cava_PRB1,darriet_JSSC}
to 4$f$ lanthanides {\it e.g.} La, Nd, Sm, Eu, and
Lu,~\cite{doi_JSSC,doi_JSSC1} to alkali and alkaline earth metals
like Li, Na, Mg, Ca, and Sr,~\cite{darriet_JSSC,loye_JACS,CaRu2_JAC}
and more.

Ba$_3M$Ru$_2$O$_9$ adopts the hexagonal BaTiO$_3$ structure (6$H$)
(Fig. 1), which consists of a pair of face-shared octahedra (blue,
Wyckoff notation: 4$f$) and a single octahedron, connecting two such
pairs (yellow, Wyckoff notation: 2$a$) through corners. Henceforth,
these two sites will be referred to as P and Q-sites, respectively.
Interestingly, it has been observed that the foreign ions ($M$)
usually show a clear affinity towards the Q-site,
~\cite{verdoes_AC,Schaller_MRB,zhao_JSSC1,donohue_IC,Battle_JSSC,Cava_JSSC,Cava_PRB1,darriet_JSSC,doi_JSSC,doi_JSSC1,loye_JACS,CaRu2_JAC}
forcing the Ru ions to occupy the P-site. In fact, it is this particular site preference of the cations that leads to the choice of the
chemical composition Ba$_3M$Ru$_2$O$_9$, where the two P-sites and the one Q-site are completely
occupied by the Ru and the $M$-ions, respectively. This naturally results in a direct Ru-Ru correlations within the two P-sites and Ru-O-\emph{M} interactions involving Ru in a P-site and the metal ion in a Q-site. One advantage of these doped compounds
is that unlike the undoped 6$H$ BaRuO$_3$, most of them could be synthesized under ambient pressure. The distance between the
Ru ions within the pair of P-sites varies between 2.5 to 2.7 \AA\, depending
on the oxidation state and ionic radius of the $M$ ion, which is
even smaller compared to the interatomic separation in pure Ru
metal. As a result of this, the magnetism of the compounds with a
nonmagnetic $M$-ion such as Ca$^{2+}$, Cd$^{2+}$, Mg$^{2+}$, In$^{3+}$, Y$^{3+}$ is mostly dominated by the Ru-Ru
antiferromagnetic interaction within the isolated Ru$_2$O$_9$ dimer
(pair of blue octahedrons in Fig. 1).~\cite{doi_JSSC,darriet_JSSC}
However, the scenario changes significantly in the case of
intervening magnetic $M$-ions such as Co$^{2+}$, Ni$^{2+}$ or Cu$^{2+}$ occupying the
Q-site, where a 3-dimensional antiferromagnetic order is realized
involving both $M$ as well as Ru ions.~\cite{Cava_JSSC,Cava_PRB1}

Among this ample variety of doped hexagonal ruthenates,
Ba$_3$FeRu$_2$O$_9$ (BFRO) is a unique member of 6H ruthenates family, due to the fact that unlike most other metals,
the Fe ion does not show such exclusive affinity towards the Q-site and can actually occupy the
P-site with relative ease.~\cite{Cava_JSSC} As a result there is
substantial Fe/Ru site disorder in this compound and consequently,
additional Ru-Fe and Ru-O-Ru configurations, other than the expected
Ru-Ru and Fe-O-Ru correlations, also get activated in this compound,
making the system magnetically more complex. However, despite of the peculiarities in this system, there are very few
experimental reports on it,~\cite{Cava_JSSC,Cava_PRB1}
and a detailed, comprehensive study is still missing.


In this paper, we report results of detailed magnetic characterization on
BFRO along with the results of long and short range structural
studies. Collectively, these results help to determine the true
structure-property relationship for BFRO. It appears that  many
competing magnetic interactions come into play in this system due to
Fe/Ru disorder, giving rise to strong magnetic frustration. As a result
of this, a global `glass'-like order sets in at low temperatures.
Interestingly, additional magnetic responses are found to exist
over a large temperature range above this glass transition temperature.
The structural information obtained from  XRD and NPD
(x-ray and neutron powder diffraction) and XAFS (x-ray absorption
fine structure) indicate that the viability of Fe and Ru ions
occupying the P and Q sites not only results in site disorder
within a unit cell but also extends much beyond
giving rise to  "clustering" with several inhomogeneous Fe and Ru rich regions.
It is observed that this extended chemical inhomogeneity play an important role in
generating magnetic metastabilities in the system much above the glass-like transition
temperature.

\section{Experimental details}
Stoichiometric amounts of BaCO$_{3}$, Fe$_{2}$O$_{3}$ and RuO$_{2}$
were ground in agate mortar and the mixture was calcined at
900$^{0}$ C for 12 h. The calcined material was heated at
1150$^{0}$ C for 72 hrs in oxygen atmosphere with several
intermediate grindings. The phase purity as well as the crystal
structure of the sample was probed by means of powder x-ray
diffraction (XRD) in a Bruker AXS: D8 Advanced x-ray diffractometer
equipped with Cu $K_{\alpha}$ radiation. The neutron powder
diffraction (NPD) measurements were carried out on powder samples
using the multi-position sensitive detector based focusing crystal
diffractometer set up by UGC-DAE Consortium for Scientific Research
Mumbai Centre at the National Facility for Neutron Beam Research
(NFNBR), Dhruva reactor, Mumbai (India) at a wavelength of 1.48
{\AA}. The samples were placed in vanadium cans that were directly
exposed to neutron beam for 300 K data. For low temperature data,
vanadium cans filled with the powder samples were loaded in a
Cryogenics make cryogen-free magnet system. XRD and NPD patterns were analyzed using Rietveld method
and the refinement of  crystal structure was carried out using the
JANA2000~\cite{jana2000} and FULLPROF~\cite{fullprof} softwares,
respectively. The magnetic properties were
studied in a Quantum Design SQUID magnetometer. The heat capacity
was measured by relaxation method in a Quantum Design PPMS.
$^{57}$Fe M\"{o}ssbauer studies were carried out on powdered samples using a
conventional constant acceleration spectrometer with a $^{57}$Co
source. The velocity calibration was
carried out using $\alpha$-Fe absorber and the typical line width
obtained in our spectrometer was 0.28mm/s. The isomer shift (IS)
reported is relative to $\alpha$-Fe. Ru \emph{K}-edge (around 22711 eV)
XAFS spectra were collected at the
GILDA-BM08 (General Italian Line for Diffraction and Absorption) beamline of the European Synchrotron radiation facility (ESRF,Grenoble, France).~\cite{GILDA} The
Ba$_3$FeRu$_2$O$_9$ ceramic pellet was initially ground in an agate
mortar using an automatic grinder in order to obtain fine powders
which were then mixed with BN powders in the grinder for several
minutes. This procedure ensures highly homogeneous samples suitable for high quality XAFS spectra. Measurements were performed
in transmission mode keeping the sample at liquid nitrogen
temperature in order to reduce thermal disorder in the structural
XAFS signal. Two scans were collected and averaged in order to
improve the signal to noise ratio. Standard procedures were adopted
for data normalization and extraction of the structural signal
$\chi(k)$ ($k=\hbar^{-1}\sqrt{2m_e (E-E_o)}\;$ ), and the
quantitative analysis of the XAFS spectrum has been performed
including single and multiple scattering terms along the lines
already described~\cite{monesi2005} using ESTRA and FITEXA programs.
The $k^2$ weighted XAFS signal $k^2 \chi(k)$ was Fourier filtered in
the $0.5\le R \le 4.2$ \AA\ range and the Filtered spectrum was
refined in the $3.5\le k \le 19$ \AA$^{-1}$ range. In the fit,
theoretical amplitude and phase functions were calculated using FEFF
8.2 code,~\cite{feff} for atomic clusters generated using the data of
XRD and NPD data refinements. In the refinement of the XAFS data,
particular care has been devoted to the next neighbour shells since
the relative multiplicity of A-B (around 2.6 \AA ) and A-O-B (around
4 \AA ) (A,B = Fe/Ru) are directly related to the local chemical
order around Ru ions. It is noticeable that RuOFe and RuORu
contributions have largely different amplitude and phase functions, therefore
giving a good chemical selectivity in the analysis.

\section{Result and discussions}
Powder XRD data collected from polycrystalline BFRO (Fig. 2(a))
confirms complete phase purity. All the peaks in the XRD pattern
could be satisfactorily indexed and refined (red curve) with 6 layered
hexagonal (6$H$) crystal structure having space group $P$6$_{3}/mmc$.
A similar result was obtained from the refinement of the powder neutron diffraction data collected at room temperature (Fig. 2(b)).
However, the initial attempt to refine the XRD pattern assuming
perfect Fe/Ru ordering i.e. complete occupancy of Fe at the 2$a$ (Q)
site and Ru in 4$f$ site (P) did not yield satisfactory results.
Consequently, each crystallographic position, 2$a$ and 4$f$, was set
as jointly occupied by Ru and Fe with constrains of total position
occupancy equal to 100\% and keeping similar ADPs (atomic displacement
parameters). This correction significantly improved the refinement,
leading to mix occupancy of both positions. In the final stage of
refinement the ADPs for each type of atoms were set to the values
extracted from NPD data and not further refined. The refinement
finally led to the formula Ba$_3$Fe$_{1.05(2)}$Ru$_{1.95(2)}$O$_9$, indicating minor loss of Ru in the sample. Neutron data
refinement resulted in similar distribution of Fe and Ru over two
crystallographic positions, however with higher standard deviations
compared to XRD. This is not surprising keeping in mind the
difference in the scattering factors for XRD and NPD. For XRD Ru
(44$e$) has almost twice the scattering power compared to Fe (26$e$),
while for NPD their scattering lengths are more close, 7.0 fm (Ru)
and 9.4 fm (Fe). The refined structural parameters and the most
significant bond lengths and angles are listed in Table I and
II, respectively. The refinement of room temperature XRD data shows
that 71\% of the Q-site is occupied by Fe atoms and the rest (29 \%) are occupied by Ru atoms. On the contrary the 83 \% of the P-sites is occupied by Ru atoms and 17 \% by Fe atoms.  The interatomic
distance between two Ru ions (Ru-Ru) in the Ru$_{2}$O$_{9}$ dimer is
2.631 \AA\  which is longer than that in Ru$^{+4}_{2}$O$_{9}$
dimer~\cite{doi_jmatchem} but shorter than that in
Ru$^{+5}_{2}$O$_{9}$ dimer.~\cite{Cava_JSSC} This indicates that the
average valency of Ru ions lies between 4 and 5 in this compound, as
is also expected from simple electron counting.


Next, we focus on the magnetic properties of this compound, which
have not been reported in detail till date. In Fig. 3(a), we show
the FC (field cooled) and ZFC (zero field cooled) magnetization
curves as a function of temperature for BFRO, measured with 100 Oe
applied field. The sample was cooled from 300 K to 2 K in absence of
field for ZFC and in presence of 100 Oe field for FC, while both ZFC
and FC data were taken during heating from 2 K to 300 K with 100 Oe
field. It is observed that the FC and ZFC curves start to diverge
from relatively higher temperatures and show a broad
peak around 140 K. However, both FC and ZFC magnetization continue
to increase with lowering temperature, and ZFC curve again exhibits
a peak around 25 K. It is important to note here that the broad peak
at the higher temperature, almost smears out in susceptibility
measurements under comparatively higher field (not shown here).
Moreover, the inverse susceptibility data measured in a magnetic
field of 5 kOe applied field (Fig. 3(b)) clearly diverges
from the standard Curie-Weiss behavior below 80 K. The high
temperature linear fitting of the inverse susceptibility yields a
$\theta$ value of -43~K, which indicates presence of
antiferromagnetic interactions in the system. Such divergence in the FC-ZFC
magnetization and also the deviation from the Curie-Weiss behavior at
higher temperature establish that certain magnetic metastabilities
develop much above the 3-dimensional
global magnetic ordering  sets in ($\le$~25 K).~\cite{Mydosh_PRB}

It is to be noted that the compounds with $M$ = Cu,
Ni, Co {\it etc.} exhibit very different low temperature magnetic
structures~\cite{Cava_PRB1} compared to their Fe-analogue. Earlier neutron diffraction studies described
long range antiferromagnetic structures~\cite{Cava_JSSC} for the Cu, Co and the Ni compounds at low temperature, while
for the Fe analogue it had been mentioned that the Fe/Ru disorder gives rise to complex
magnetic interactions. For example, presence of a variety of local magnetic configurations~\cite{Cava_PRB1}
and absence of any magnetic ordering in the neutron diffraction measurements~\cite{Cava_JSSC} have indeed been
discussed before. Therefore, the magnetic structure of BFRO develops differently as a result
of the Fe/Ru site disorder. In order to confirm the nature of the
magnetic order at low temperatures, isothermal remanent
magnetization (IRM) measurements have been carried out on BFRO. For
this purpose, each time the sample was cooled from 300 K in zero
field to the measuring temperature, 5 kOe field was applied for 5
minutes, and then \emph{M} was noted as a function of time
(\emph{t}) immediately after the field was switched off. The results
are shown in Fig. 3(c). For \emph{T}= 1.8 K and 15 K,
\emph{M$_{IRM}$} undergoes a slow decay, indicative of a glassy
behavior in the system. The IRM at these two temperatures can be
fitted well with the logarithmic function,
\emph{M$_{IRM}$}=\emph{M$_{0}$}-\emph{S}ln(1+\emph{t}/\emph{t$_{0}$}),
as shown by red solid lines in Fig. 3(c). The logarithmic time
dependence of the IRM is observed in magnetic materials with
hysteretic magnetization and/or glassy
systems.~\cite{magneticrelaxation_EVS,Nd2PtSi3_ssc} Generally,
materials with high coercivity show a pronounced time dependent
behavior, however, if the applied field is higher than the coercive
field the relaxation is attributed to glass-like behavior. Here, at
1.8 K the coercive field is 1140 Oe (Fig. 3(d)), whereas the
relaxation measurements were performed after the application of 5
kOe field, which is much higher than the coercive field. Hence, the
strong relaxation effects in this case are obviously of microscopic
rather than macroscopic (domains) in origin.~\cite{Ho5SiGe_PRB} The
values of the viscosity coefficient \emph{S} are 0.0037 emu/mole and
0.0067 emu/mole for 1.8 K and 15 K, respectively. These IRM measurements indicate
that unlike the Cu, Co, or Ni compounds, Ba$_3$FeRu$_2$O$_9$ undergoes
a `glass'-like transition with lowering temperature.

In panel 3(d), the ZFC as well as FC (cooled under 50 kOe applied
field) \emph{M} vs. \emph{H} loops of BFRO at 2 K are shown. The ZFC
$M$($H$) curve shows a clear hysteresis loop with coercivity of 1140
Oe but without any signature of saturation till the highest field of
measurement. Interestingly, such loops are not uncommon among spin
glasses {\it e.g.} the well known canonical spin glass alloys, CuMn
or AgMn, also exhibit identical hysteresis loop at low
temperatures.~\cite{kouvel_61_JPCS} The absolute value of \emph{M}
at 50 kOe magnetic field is found to be only 0.51 $\mu$$_{B}$/f.u. (formula
units), indicating presence of a significant number of uncorrelated spins.
However, the most fascinating
feature of this measurement is the large shift of the low
temperature $M$($H$) loop collected under field-cooled condition.
However, observation of such `exchange bias' effect is also common in the
well known spin glass alloys, which arises due to the coexisting ferro-
and antiferromagnetic interactions between the magnetic ions, placed
at different separations because of the intrinsic composition fluctuation.~\cite{kouvel_60_JAP,kouvel_61_JPCS,kouvel_63_JPCS,kouvel_NiMn_59_JPCS,kouvel_84_JAP,Wu_91_JAP}
Therefore, similar spatially varying chemical fluctuations might exist also in BFRO, which results in such resemblances with the magnetic behaviors of other canonical spin glass metallic alloy systems.

Identification of the low temperature magnetic transition as a transition to a `glass'-like magnetic phase is further realized by the temperature dependent heat capacity (\emph{C}) measurements. It is well known that spin glass transitions do not produce any sharp feature in the \emph{C} {\it vs.} $T$ measurement,~\cite{Wenger_C_PRB,Uemara_C_PRB,EVS_C_PRB1,EVS_C_PRB2} because above a few Kelvin temperature, the non-magnetic contributions to $C$ seems to overwhelm the magnetic or spin-glass terms. However, there are suggestions of extracting the signature of the `freezing' phenomena from the \emph{C/T} (=d$S$/d$T$) vs. \emph{T} plots instead,~\cite{Mydosh_book} but even in such attempts only a very weak feature could be observed near the freezing temperature. In the present case, it is noticed that \emph{C} gradually decreases as \emph{T} decreases down to 1.8 K
without any evidence for any well defined peak characterizing a long
range magnetic order, as shown in Fig. 4(a). However, when the data
is plotted in the form of \emph{C/T} vs. \emph{T}, a very weak feature
could be seen at $\sim$25 K as indicated by an arrow in Fig. 4(b). The
small size of this feature indicates that the entropy change
associated with the transition is very small, as expected in a glassy transition with randomness in magnetic interactions. Further, the NPD data at 2 K (Fig.4(c)) exhibit neither any significant change in any peak intensity nor
the development of any additional peaks compared to those at room temperature,
signifying the fact that at 2 K, the magnetic
phase is neither ferromagnetic nor antiferromagnetic (within the
resolution of the instrument). The spectral pattern is modified at
certain angles due to the interference from the magnet shroud and
those 2$\theta$ regions are excluded from refinement. In this
refinement, the occupancies of Ru and Fe at the 2$a$ and 4$f$ sites
were kept fixed at the values obtained from the refinement of room
temperature NPD data. All peaks of the observed pattern were
reproducible in the refinement with space group $P$6$_{3}/mmc$
without any significant magnetic contribution. This result clearly confirms that there is no structural
transition at low temperature and also the magnetic transition at 25 K does not led to any long
range magnetic ordering. Therefore, the heat capacity and the NPD results presented in Fig. 4 with the description that the low temperature
magnetic transition observed in this system is probably a glassy one.

Now, it has been shown above (Fig. 3) that magnetic metastabilities exist much above the glass transition temperature. Therefore, next we concentrated on measurements above 25 K in order to understand the high temperature magnetic structure of the system. The \emph{M} vs. \emph{H} were measured at different temperatures and are shown in Fig. 5(a). For better visual clarity
we have plotted only the first quadrant of the loop and to focus on
the remanent magnetization, we have plotted the same data in Fig
5(b) in an expanded scale. For each temperature, the loop has two
branches, one from 0 Oe to 50 kOe and other from 50 kOe to 0
Oe as indicated by arrows for 1.8 K data in Fig. 5(b). At 1.8 K,
there is a remanent magnetization of 0.015 $\mu$$_{B}$/f.u., which
lowers down to 0.007 $\mu$$_{B}$/f.u. at 15 K. The hysteresis loops
measured at 30 K and higher hardly show any remanent magnetization,
indicating near hysteresis loss above 25 K. However, IRM
measurement at 80 K exhibits a small remanence which
remained almost constant with time, as shown in Fig. 5(c). The
magnitude of this constant IRM at 80 K is approximately 0.0425
emu/mole which is small but not absolutely negligible considering the fact that
the value of IRM is only 0.1375 emu/mole at t=0 at 15 K. Therefore,
the higher temperature magnetic data reveal contradictory behaviors
such as weak but stable remanence as well as a clear deviation from
the Curie-Weiss behavior below 80 K even though loss of perceptible hysteresis in \emph{M}(\emph{H})
curves occurs above 25 K. Again, it is worth mentioning here that such
behavior, much above the spin glass transition temperature, is
indeed observed in many canonical spin glass systems like the well
known CuMn or AgMn metallic
alloys,~\cite{kouvel_60_JAP,kouvel_61_JPCS,kouvel_63_JPCS} where the
statistical compositional fluctuations inherent to atomically
disordered systems have been held responsible for such magnetic
features. Therefore, the observed anomalous behaviors are definite indications
of many local magnetic configurations above 25 K in this case,~\cite{Cava_PRB1} which might appear
due to chemical inhomogeneities in the system. The origin of such inhomogeneities could be the relative distribution of Fe and
Ru ions in the compound, as is revealed by XAFS measurements, discussed later.


However, in order to probe the presence of such inhomogeneous magnetic
ordering, we have also carried out $^{57}$Fe M\"{o}ssbauer experiments at
different temperatures, shown in Fig. 6. The room temperature
spectrum consists of a pure quadrupolar split (e$^{2}$qQ/2 = 0.3
mm/s) spectral line which indicates  paramagnetic nature of the
sample at room temperature. It shows an isomer shift of 0.43 mm/s
with respect to Fe metal (typical of a high spin Fe$^{3+}$ species) and it undergoes thermal red shift with
decreasing temperature. However, at sufficiently low temperatures
there is a clear decrease in intensity of the absorption line
accompanied by a broadening of the line signifying the appearance of
magnetic hyperfine field (see the line at 48 K or below). This loss
of paramagnetic intensity could be interpreted by the formation of
broad hyperfine split signal, appearing from locally ordered Fe ions
at the cost of paramagnetic, uncorrelated Fe spins. As the loss in
the paramagnetic intensity occurs gradually with lowering
temperature, there is obviously no sharp para- to ferro- transition,
reminiscent of a series of magnetic transitions. Finally, below 30
K, the paramagnetic line nearly disappears and a well resolved
hyperfine split spectra typical of magnetically ordered systems
appears, confirming the onset of a magnetic transition below this
temperature. The magnetic hyperfine field at 4.2 K is found to be
$\sim$ 49 T, which is similar to that observed in Fe$_{3}$O$_{4}$.
It should be noted that even at 4.2 K, a small paramagnetic
component ($\sim$ 8\%) is observed which may imply that some Fe ions
are yet to order at this temperature. All of these observations are
consistent with the inferences obtained from the
magnetization studies.

Finally, we performed XAFS studies to probe the origin of
inhomogeneous magnetic structure at higher temperature. It is likely that the Fe/Ru disorder can extend further, giving rise to spatially extended regions with different Fe neighborhoods, ranging from isolated Fe atoms to
extended Fe ``clusters" with Fe filling both the nearest neighbor P and Q sites. Now, such
different regions can order at different temperatures depending on their
size and composition. Although the diffraction probes undoubtedly
confirmed the Fe/Ru site disorder within an unit cell, such measurements cannot reveal
different local compositions, which can be substantially different and would
actually manipulate the physical properties heavily.~\cite{xafs_PRL}
The analysis of Ru \emph{K}-edge XAFS data allows us to achieve detailed insight about the local chemical structure on BFRO.
 Fig. 7(a) and 7(b) show the Ru $K$-edge XAFS data, and its Fourier transform along
with the respective best theoretical curves. The structural parameters
obtained from XAFS analysis are presented in Table III. The most
important point to note here is the fact that Ru finds higher number
of Ru neighbors both in the P and Q-sites, compared to
the findings of the diffraction experiments. It is to be noted that
unlike XAFS, the inherent assumption for XRD or NPD refinements is
that the overall chemical composition of a material must be
maintained within the single unit cell. On the other hand, XAFS technique being a local probe, can  provide more detailed insight on the relative arrangement of Fe and Ru ions without any such restriction.\cite{xafs_PRL} In the first attempt, the XAFS data were refined constraining the P and Q site occupancies to the values
 obtained by diffraction data. However the quality of the
 XAFS data refinements definitely improves (the $R^2$ factor is sensibly reduced) on removing these
constraints. In this way, the analysis of the Ru local environment depicts noticeable chemical
inhomogeneity beyond an unit cell volume, where clear Ru-rich
regions are found to be present in the system. The much larger Ru-Ru
(between P-sites around 2.6 \AA ) and Ru-O-Ru (linking P-Q sites,
around 3.9 \AA) connectivities, observed in XAFS compared to the
same from XRD (shown within square brackets in Table III)
conclusively establish this fact. It is to be noted that Ru-O-Fe and
Ru-O-Ru distances are found very similar within our experimental
accuracy and might have certain errors in quantification. However, as the XAFS results clearly indicate presence of Ru-rich
areas, Fe-rich regions also have to be formed in the system.
Therefore, it appears that the ability of the Fe ions to fit in both
the P and Q-sites not only creates Fe/Ru site disorder within a unit
cell but also allows preferential clustering to occur over certain
spatial domains. Obviously, such a possibility does not exist for
other Ba$_3M$Ru$_2$O$_9$ compounds, where the occupancy of the $M$
ions is always restricted to Q-site and no $M$-$M$ or
$M$-O-$M$ correlations are permitted. Overall, our long range and
local structural investigations prove that the microscopic chemical
composition can be vastly different from assumed long range
structures and also show that the Fe-member of Ba$_3M$Ru$_2$O$_9$
series of compounds clearly stands out and altogether belongs to a
different class with respect to other members of the family.

Now, it is possible to correlate the observed chemical
`inhomogeneity' with these anomalous magnetic behaviors for \emph{T} $>$
25K. It can be easily speculated that different Fe-rich regions of
various spatial extents and compositions may start to order magnetically at higher
temperatures, while they remain uncorrelated with respect to each
other. The $M$($H$) curve of such a mixture fails to exhibit
distinct loops but anomalies at higher temperature become visible in
IRM or inverse susceptibility behaviors. However, when the
temperature is lowered to 25 K, a global magnetic correlation sets
in and many different possible interactions give rise to strong
frustration and a 3-dimensional `glassy' order.

\section{Conclusion}
In summary, our detailed experimental study establishes
Ba$_3$FeRu$_2$O$_9$ compound to be a special one with respect
to other members of the family, where the possibility of Fe/Ru-site
disorder gives rise to many competing magnetic interactions and as a
result, a global `glass'-like ordering occurs at lower temperature. The random distribution of Fe and Ru ions also gives rise
to spatial inhomogeneity that further complicates the magnetic
structure. The Fe-rich regions start to order at rather higher
temperature and probably, a distribution of magnetic transitions exist at higher temperatures.

\section{Acknowledgement}
SM thanks CSIR, India for fellowship. SR thanks UGC-DAE CSR Mumbai
Centre, DST-RFBR and DST Fast Track, India for financial support.
DDS acknowledges J. C. Bose National Fellowship.

\newpage

\begin{table}
\caption
 {Structural parameters for Ba$_{3}$FeRu$_{2}$O$_{9}$}

\vspace*{0.1in} The atomic positions: Ba(1): 2b(0, 0, 1/4); Ba(2):
4$f$(1/3, 2/3, z); Ru/Fe: 4$f$(1/3, 2/3, z); Fe/Ru: 2$a$(0, 0, 0);
O1: 6h (x, 2x, 1/4); O2: 12k(x, 2x, z)

Constrained: B$_{Ba(1)}$=B$_{Ba(2)}$

\begin{tabular}{l c c }
\hline
\hline
Atom & Parameter & XRD at 300 K \\
\hline
  & a (\AA) & 5.7310(2)\\
  & c (\AA) & 14.0768(8)\\
  & V (\AA$^{3}$) & 400.396(3) \\
  Ba1 & B (\AA$^{2}$)& 0.5053 \\
Ba2 & z & 0.0908(1)\\
   & B (\AA$^{2}$)&0.5053\\
Ru/Fe (P-site) & z & 0.8434(1)\\
      & B (\AA$^{2})$& 0.3474\\
    & n & 0.833(8)/0.167(8) \\
Fe/Ru (Q-site) & B (\AA$^{2}$)&0.3474 \\
      & n & 0.71(1)/0.29(1) \\
O(1)  & x & 0.508(1) \\
      & B (\AA$^{2}$)& 0.4343\\
O(2)  & x & 0.837(1) \\
      & z & 0.0811(5)\\
      & B (\AA$^{2}$)& 0.4343\\
      & R$_{p}$(\%)& 9.98\\
      & R$_{wp}$(\%)&15.35\\
      & $\chi$$^{2}$&1.26\\
\hline
 \end{tabular}
\end{table}

\newpage

\begin{table}
\caption
 {Selected bond lengths (\AA) and angles ($^{0}$)  for Ba$_{3}$FeRu$_{2}$O$_{9}$}

\vspace*{0.1in}

\begin{tabular}{l c  c }

\hline
\hline
  & & XRD 300 K \\
\hline
Ba1-O1    $\times$ 6& & 2.865(7)\\
Ba1-O2    $\times$ 6& & 2.879(9)\\
Ba2-O1    $\times$ 3& & 2.831(8)\\
Ba2-O2    $\times$ 6& & 2.869(9) \\
Ba2-O2    $\times$ 3& &2.949(6)\\
Ru/Fe-O2  $\times$ 3& & 1.993(9)\\
Ru/Fe-O1  $\times$ 3& & 2.056(9)\\
Fe/Ru-O2  $\times$ 6& & 1.984(8)\\
Ru/Fe-O (average)   & & 2.025\\
Ru/Fe-Ru/Fe         & & 2.631(2)\\
$\angle$Ru/Fe-O1-Ru/Fe      & & 79.55(6) \\
$\angle$Ru/Fe-O2-Fe/Ru      & & 177.09(9) \\

 \hline

 \hline
 \end{tabular}

\end{table}

\newpage

\begin{table}
\caption {Structural results obtained from the refinement of the Ru
K edge XAFS spectrum. The multiplicity numbers  of Ru-Ru(Fe) and
Ru-O-Ru(Fe) reported within square brackets are the calculated
numbers using the information of Fe/Ru occupancies at P and Q site
obtained from XRD. The $R^2$ value written within the square
brackets is  obtained with the constraints imposed by random
distribution of Fe/Ru on P and Q sites. In numbers within brackets
are the errors on the last digit of the refined parameters.}
\begin{tabular}{lccc}
\hline
\hline
Shell   & N & R (\AA) & $\sigma^2 (\times 10^3$\AA$^2$) \\
\hline
RuO    & 6                & 1.976(6)  & 2.9(2)\\
RuRu   & 0.89(4) [0.69]   & 2.61(1)   & 4.5(6)\\
RuFe   & 0.11    [0.31]   & 2.62(1)   & 5.5(4)\\
RuBa   & 7.6(2)           & 3.49(1)   & 7.2(4)\\
RuORu  & 2.4(1) [0.91(8)]  & 3.91(2)   & 3.3(3))\\
RuOFe  & 1.9(1) [2.33(2)]  & 3.92(2)   & 1.6(2)\\
ORuO   & 6                & 3.90(2)   & 3.6(9)\\
RuO    & 20(2)            & 4.56(2)   & 15.(5)\\
\hline
\multicolumn{3}{l}{R$^2$ = 0.93\% [=1.1\%]}\\
\end{tabular}
\end{table}

\clearpage

\begin{center}
\begin{figure}
\resizebox{8cm}{!}
{\includegraphics*[3mm,50mm][165mm,185mm]{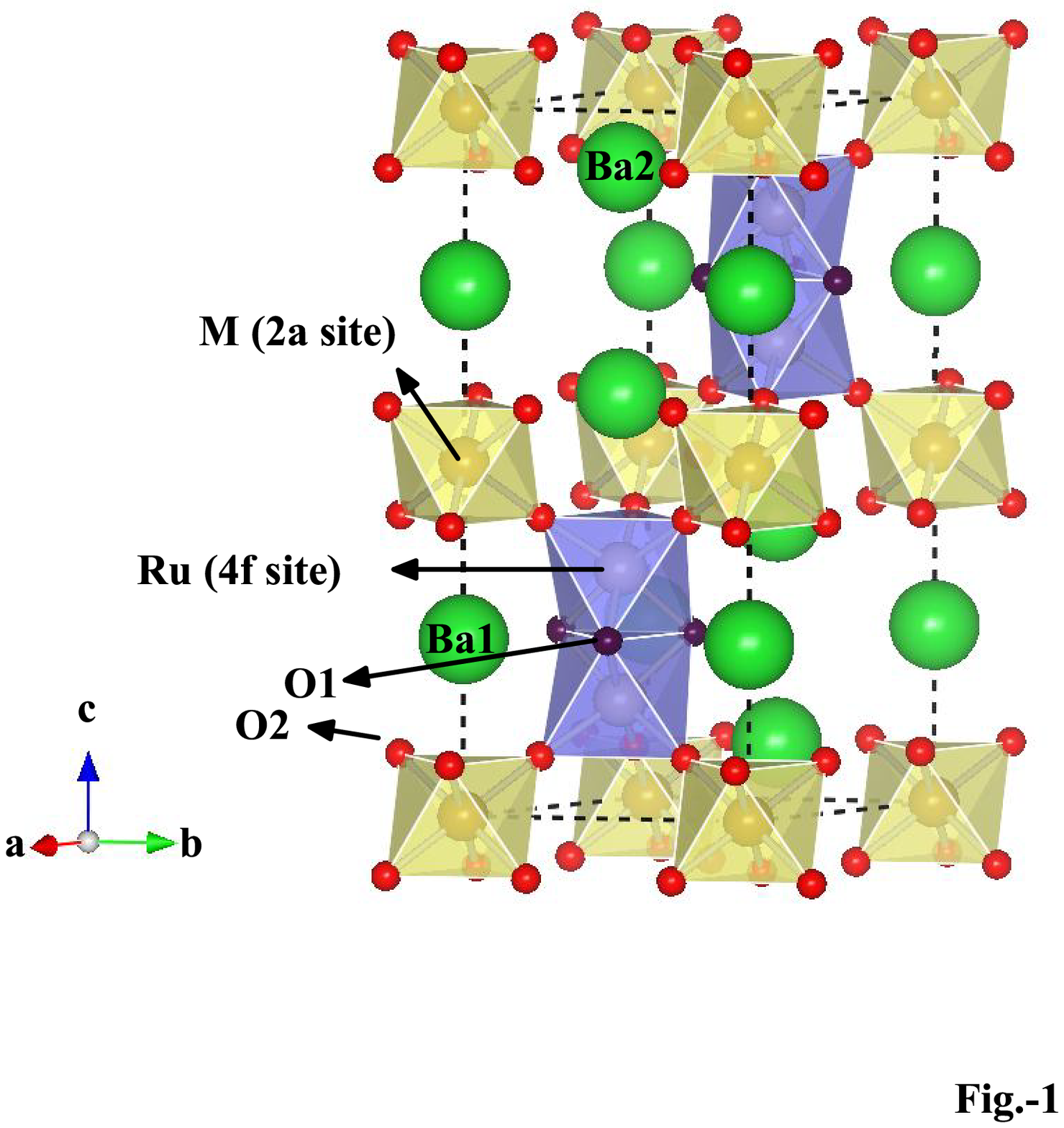}} \\
\caption{(color online) The ordered crystal structure of Ba$_{3}$MRu$_{2}$O$_{9}$.}
\end{figure}
\end{center}

\begin{center}
\begin{figure}
\resizebox{8cm}{!}
{\includegraphics*[6mm,10mm][185mm,250mm]{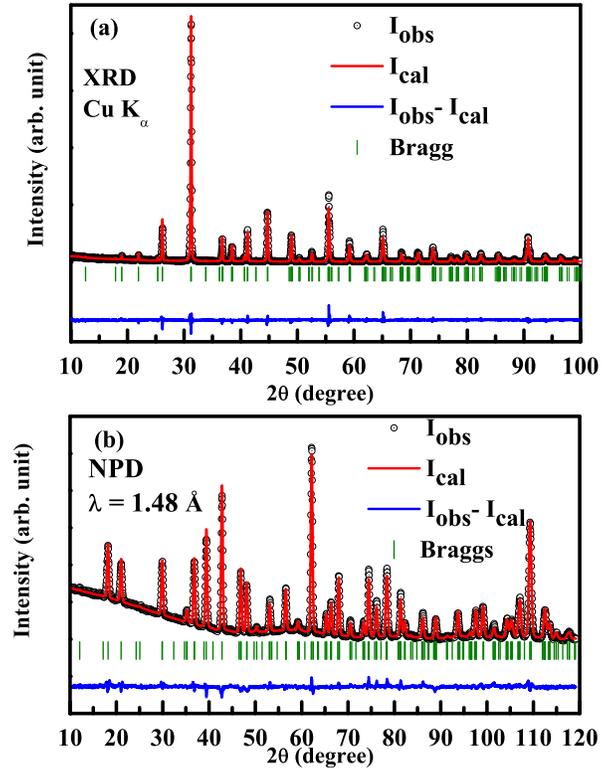}} \\
\caption{(color online) Experimental (black) and refined (red) XRD
patterns (a) and neutron powder diffraction patterns (b)
for Ba$_{3}$FeRu$_{2}$O$_{9}$ at room temperature.}
\end{figure}
\end{center}

\begin{center}
\begin{figure}
\resizebox{8cm}{!}
{\includegraphics*[3mm,10mm][195mm,275mm]{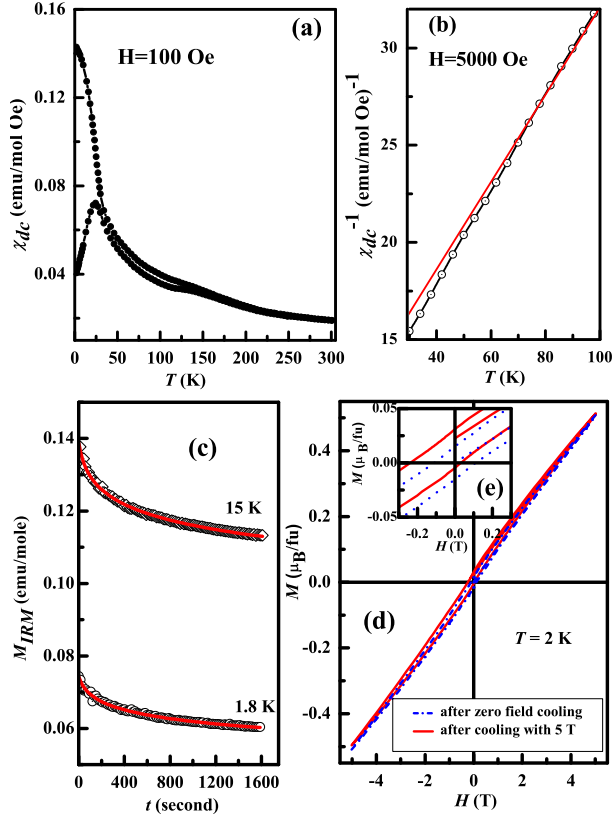}} \\
\caption{(color online) Field cooled and zero field cooled
magnetization curve vs. \emph{T} under \emph{H} = 100 Oe is shown in
(a). The inverse susceptibility as a function of temperature
for \emph{H} = 5 kOe is shown in (b), while the solid line
expresses the Curie-Weiss behavior. The deviation from the
Curie-Weiss nature is observed below 80 K. The relaxation of IRM as
a function of time at selected temperatures are shown in (c).
The ZFC as well as FC (cooled under 50 kOe applied field) \emph{M}
vs. \emph{H} loops at 2 K are shown in (d), while an expanded
view is shown in (e).}
\end{figure}
\end{center}

\begin{center}
\begin{figure}
\resizebox{8cm}{!}
{\includegraphics*[8mm,10mm][199mm,275mm]{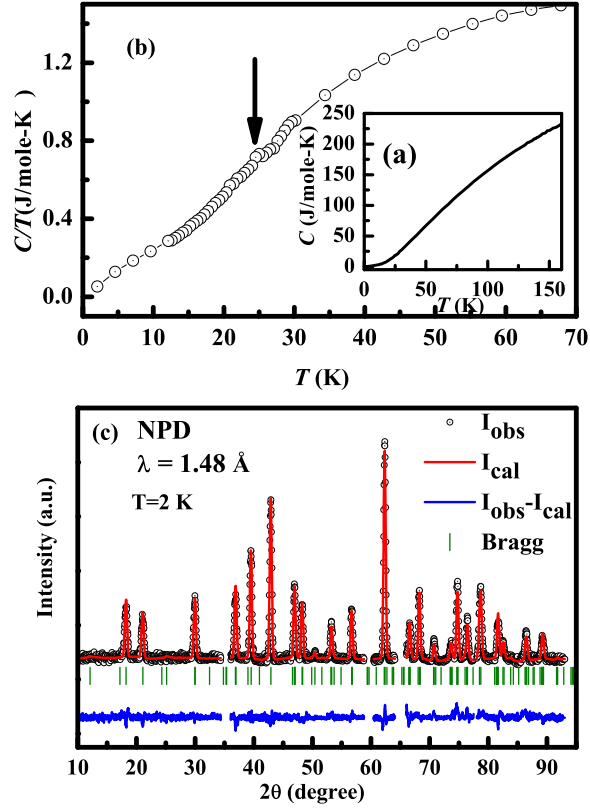}} \\
\caption{(color online) The variation of heat capacity {\emph{C}}
with temperature (\emph{T}) is shown in (a) while \emph{C/T}
vs. \emph{T } is plotted in (b). (c) shows the
experimental (black) and refined neutron powder diffraction patterns
for Ba$_{3}$FeRu$_{2}$O$_{9}$ collected at 2 K.}
\end{figure}
\end{center}

\begin{center}
\begin{figure}
\resizebox{8cm}{!}
{\includegraphics*[3mm,20mm][185mm,280mm]{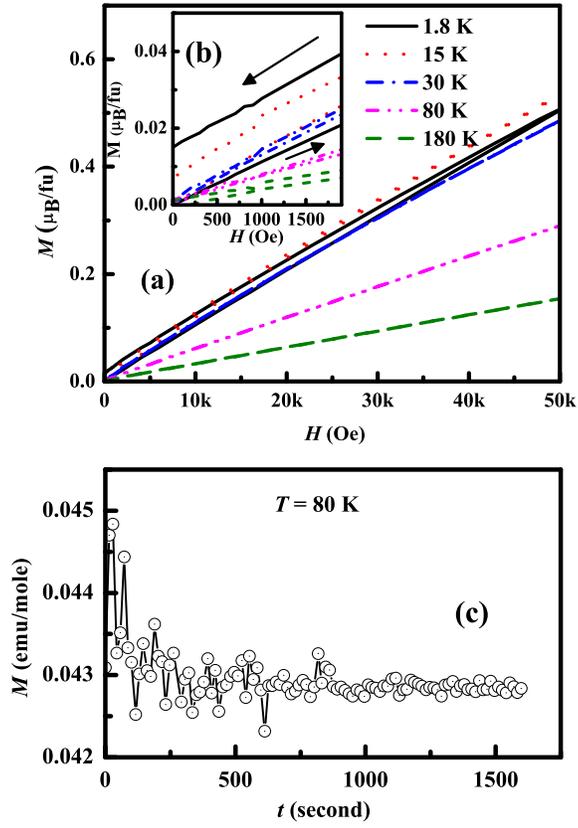}} \\
\caption{(color online) The \emph{M} vs. \emph{H} at different
temperatures are shown in (a), while an expanded view near
zero field are plotted in (b). (c) shows the variation
of IRM with time (\emph{t}) at 80 K.}
\end{figure}
\end{center}

\begin{center}
\begin{figure}
\resizebox{8cm}{!}
{\includegraphics*[10mm,40mm][170mm,250mm]{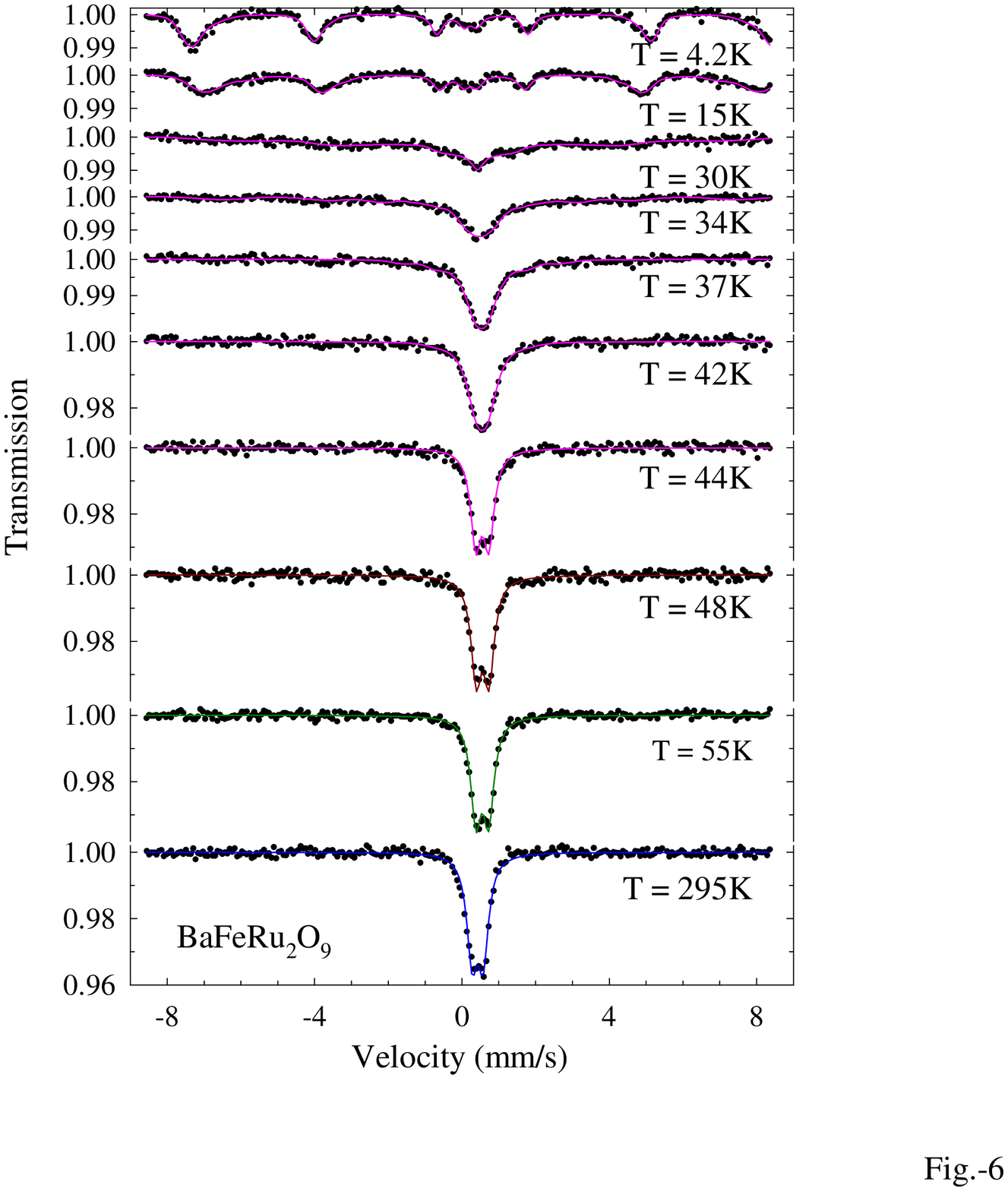}} \\
\caption{(color online) $^{57}$Fe M\"{o}ssbauer data collected at different temperatures.}
\end{figure}
\end{center}

\begin{center}
\begin{figure}
\resizebox{8cm}{!}
{\includegraphics*[18mm,18mm][143mm,105mm]{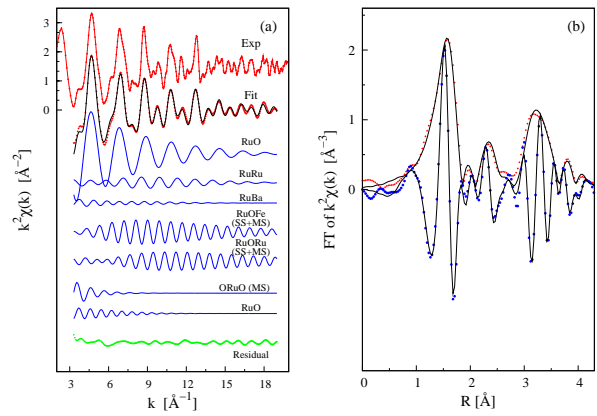}} \\
\caption{(color online) (a) and (b) show the Ru $K$-edge XAFS data, and its Fourier transform along with the respective best fit spectra.}
\end{figure}
\end{center}

\end{document}